# Enhanced exciton-exciton collisions in an ultra-flat monolayer MoSe$_2$ prepared through deterministic flattening

T. Hotta[1], A. Ueda[1], S. Higuchi[1], M. Okada[2], T. Shimizu[2], T. Kubo[2], K. Ueno[3], T. Taniguchi[4], K. Watanabe[5] and R. Kitaura[1,*]

[1]Department of Chemistry, Nagoya University, Nagoya, Aichi 464-8602 Japan
[2]Nanomaterials Research Institute, National Institute of Advanced Industrial Science and Technology (AIST), Tsukuba, Ibaraki 305-8565, Japan
[3]Department of Chemistry, Saitama University, Saitama 338-8570, Japan
[4]International Center for Materials Nanoarchitectonics, National Institute for Materials Science, Tsukuba, Ibaraki 305-0044, Japan
[5]Research Center for Functional Materials, National Institute for Materials Science, Tsukuba, Ibaraki 305-0044, Japan

Corresponding Author: R. Kitaura, r.kitaura@nagoya-u.jp

Abstract
Squeezing bubbles and impurities out of interlayer spaces by applying force through a few-layer graphene capping layer leads to van der Waals heterostructures with ultra-flat structure free from random electrostatic potential arising from charged impurities. Without the graphene capping layer, a squeezing process with an AFM tip induces applied-force-dependent charges of $\Delta n \sim 2 \times 10^{12}$ cm$^{-2}$ μN$^{-1}$, resulting in strong intensity of trions in photoluminescence spectra of MoSe$_2$ at low temperature. We found that a hBN/MoSe$_2$/hBN prepared with the "modified nano-squeezing method" shows a strong excitonic emission with negligible trion peak, and the residual linewidth of the exciton peak is only 2.2 meV, which is comparable to the homogeneous limit. Furthermore, in this high-quality sample, we found that formation of biexciton occurs even at extremely low excitation power ($\Phi_{ph} \sim 2.3 \times 10^{19}$ cm$^{-2}$ s$^{-1}$) due to the enhanced collisions between excitons.

Keywords: two-dimensional materials, van der Waals heterostructures, biexcitons

Introduction

2D materials, such as graphene and transition metal dichalcogenides (TMDs), have provided fascinating fields for investigation of physics in the realm of two-dimensional systems[1-5]. The reduced dimensionality in 2D materials lead to significant many-body effects due to the weaken Coulomb screening and confinement of carriers, resulting in strong excitonic effects in response to optical excitations[6-9], enhanced charge-density wave order[10-13], and etc. The reduced dimensionality also leads to broken inversion symmetry, which causes emergence of the valley degree of freedom in, for example, a 2D TMD[14-17]. Moreover, the Fermi level of 2D materials can easily be tuned through applying gate voltage in 2D-material-based field-effect transistors[18-21]. These excellent characteristics of 2D materials enable us to explore a wide range of basic physics and possibilities on realization of novel optoelectronic devices.[3, 22]

For exploration of intrinsic properties of 2D materials, the environmental effects need to be suppressed. This is because, in the case of atomically thin materials, environments, including substrates and adsorbates, can strongly alter their optical and electronic properties. For example, substrate-induced roughness causes inhomogeneity in electronic structure and additional carrier scatterings, leading to inhomogeneous broadening in optical spectra and reduction of carrier mobility in 2D materials[23-25]. Furthermore, substrates sometimes have charged impurities and low-energy phonons, which also contribute to reduction of carrier mobility. In fact, early works on graphene on $SiO_2$/Si shows mobility $\sim 10^4 cm^2/Vs$[26] even at cryogenic temperature, which is far from the theoretical phonon-limit carrier mobility[27]. Adsorbates also cause inhomogeneous broadening in optical spectra and reduction of carrier mobility in a similar way. Suppression of theses unwanted environmental effects thus crucial for investigations of intrinsic properties of 2D materials.

One of the best ways to suppress the environmental effects are encapsulating 2D materials with hexagonal boron nitride (hBN) flakes: hBN/a 2D layer/hBN. hBN is a layered insulator with large bandgap of $\sim 6$ eV[28]. The important thing is that hBN has an atomically flat surface without dangling bonds, charged impurities and low-energy optical phonons. 2D materials encapsulated by hBN flakes, therefore, are free from the environmental effects arising from substrates and adsorbates. For example, in hBN-encapsulated semiconducting TMDs, full width at half maximum (FWHM) of photoluminescence (PL) peaks is much smaller than those of TMDs on $SiO_2$/Si, reaching to the value of the intrinsic limit (neutral exciton of $MoS_2$: $\sim 2$ meV at 4 K)[24]. Sharp optical responses in hBN-encapsulated structures have led to observations of various intriguing optical responses originating from long-lived interlayer excitons and moiré excitons etc[29-31].

Whereas hBN-encapsulated structures give the ideal platform for 2D materials research, its fabrication process is one of the most serious bottlenecks in the research on 2D materials. The standard technique for the fabrication is the dry-transfer method based on the pick-up-and-drop process, where 2D layers are consecutively picked up with the top hBN flake. In this process, bubbles or impurities derived from hydrocarbon are inevitably encapsulated between layers[32]. The bubbles and impurities are sources of inhomogeneity in electronic structure and carrier scatterings, significantly damaging the quality of hBN-encapsulated structures. Performing the process under vacuum condition or repeated pressing against a substrate[33, 34] can reduce the amount of bubbles and impurities, but complete removal is still very difficult. If we can develop a sure way to fabricate high-quality atomically-flat vdW heterostructures, it should greatly contribute to exploration of fascinating possibilities of 2D materials.

In this work, we have developed a new method, the modified nano-squeezing method, to fabricate high-quality ultra-flat vdW heterostructures for observation of their intrinsic properties. Sweeping the sample surface with an AFM tip, the nano-squeezing method[35, 36], is one of the sure ways to remove impurities and bubbles encapsulated between layers in vdW heterostructure (Fig. 1a). We found, however, that optical responses and electronic transport properties of samples prepared by the nano-squeezing method is strongly influenced by charged impurities: intense PL from trions and saturation of carrier mobility at cryogenic temperature. This strongly suggests that the nano-squeezing process cause significant amount of charged impurities on the surface of vdW heterostructures. To suppress the effect from charged impurities, we put a few-layer graphene as a capping layer before the squeezing process; the graphene capping layer can effectively screen the unwanted effect from charged impurities attached (Fig. 1b). PL spectra of the hBN/MoSe$_2$/hBN prepared with the modified nano-squeezing method gives very low intensity of trion peak even at cryogenic temperature and a very narrow linewidth approaching the intrinsic limit. Moreover, we found that collisions between two excitons occurred even at very low excitation power with CW laser (photon flux: $\Phi_{ph}$ ~ 2.3 × 10$^{19}$ cm$^{-2}$ s$^{-1}$) in the high-quality sample prepared with the modified nano-squeezing method.

Results and discussion

We fabricated hBN/MoSe$_2$/hBN by the standard dry transfer method (see experimental and supplement information). Figure 2a-c shows an optical microscope image of a hBN/MoSe$_2$/hBN sample before and after the nano-squeezing process. As seen in the images, black contrasts arising from encapsulated bubbles and impurities disappears after the nano-squeezing process at 0.8 μN; bubbles and impurities are removed to be accumulated at the edges of the squeezed area. Residual mean square (RMS) roughness at the squeezed area (Fig. 2c) is only 0.36 nm, clearly showing atomically flat surface after the squeezing process.

Figure 3a shows PL spectra at 10 K before and after the squeezing. Two sharp peaks at ~ 1.64

eV and ~ 1.61 eV arise from radiative recombination of excitons and trions, respectively[37]. In the case of non-squeezed (dirty) sample, the additional broad peaks at 1.50 eV, which originates from bound excitons trapped around impurities[38], are observed. On the other hand, in the case of the squeezed sample, there are no peaks at low energy region, and the linewidth of exciton emission is reduced from 12 to 4 meV. These results clearly demonstrate that inhomogeneous broadening in PL spectra can be strongly suppressed by the squeezing process, being consistent to the reduction in RMS roughness after the squeezing. The ultraflat structure obtained with the squeezing method is essential to observe sharp optical responses of 2D materials.

Although the encapsulated impurities are removed by the squeezing process, low-temperature PL spectroscopy shows a significant effect of charged impurities. Figure 3a shows PL spectra of squeezed samples measured at 10 K. As clearly seen, PL spectra significantly change depending on the force applied during the squeezing process; the intensities of low energy peak increase as the force increases. The low-energy peak is assigned to radiative recombinations of trions because of the binding energy of 30 meV[39], excitation power dependence of peak intensity[40-42] and gate-voltage dependence of PL spectra[43] (Fig. S2b). As shown in the Fig. 3b, intensity ratio between the exciton peak and the trion peak ($I_{ex}/I_{tr}$) shows exponential decrease against the force applied, and similar exponential decrease in $I_{ex}/I_{tr}$ is also seen in the gate voltage dependence. This means that the stronger the force applied during a squeezing process is the more carrier is induced in the encapsulated monolayer $MoSe_2$. Based on comparison between the force dependence and the gate-voltage dependence, we estimated that the amount of tip-induced carriers ($\Delta n$) as $\Delta n \sim 2 \times 10^{12}$ cm$^{-2}$ μN$^{-1}$ (details are shown in Fig. S3). Because an AFM tip directly scratches the surface of a top hBN flake during a squeezing process, the process can etch top most hBN surface to induce dangling bonds or cause impurities arising from the AFM tip. These dangling bonds and the tip-induced impurities can be charged impurities sitting adjacent to the encapsulated $MoSe_2$, and the induced charged impurities probably lead to the observed decrease in $I_{ex}/I_{tr}$ shown in Fig. 3b.

To confirm the existence of charged impurity after a squeezing process, we measured temperature dependence in carrier mobility of a hBN/graphene/hBN sample, which was prepared through the standard pick-up-and-drop method and the squeeze method with applied force of 1.8 μN. Figure 3c represents temperature dependence of electron and hole mobility, $\mu_e$ and $\mu_h$, determined from hall measurements performed in the range of 4.5 to 300 K (Fig. S4). As show in the Fig. 3c, $\mu_e$ and $\mu_h$ increase as temperature decreases. In the high temperature region, the observed relation between mobilities and temperature is close to $\mu_e$ ($\mu_h$) $\propto T^{-1}$, which is consistent to acoustic phonon scattering in two-dimensional electronic systems[27]. In contrast, the increase in mobilities deviates from the relation, $\mu_e$ ($\mu_h$) $\propto T^{-1}$, as temperature decrease and saturated below 80 K to be ~ 2.7 × 10$^4$ ($\mu_e$) and 2.2 × 10$^4$ ($\mu_h$) cm$^2$ V$^{-1}$ s$^{-1}$. Surface roughness of this sample is small enough (RMS roughness is 0.18 nm) after the squeezing process, and the roughness scattering in this sample should be strongly suppressed. The observed leveling off in carrier mobilities thus should arise from charged impurities scattering[44], being consistent to the results of PL spectroscopy obtained with a hBN-encapsulated monolayer $MoSe_2$. Dirac points in output curves of the hBN/graphene/hBN (Fig. S4d) does not locate at zero, indicating that carriers are induced in response to charges induced on the top of hBN after the process.

In order to further confirm if charged impurities can be induced by squeezing processes or not, electrostatic force microscope (EFM) observations were performed. Figure 4b-d show EFM images and the corresponding line profiles of a hBN/hBN flake (Fig. 4a) before and after a nano-squeezing process. As clearly seen in these images, electrostatic potential at the hBN surface significantly changes after the squeezing process (an increase from 0.5 to 0.7 mV). Note that morphology and height of the hBN flake remain intact after the squeezing process, and the observed change in electrostatic potential should be caused by the change in surface properties. The increase in electrostatic potential, therefore, strongly indicate that squeezing processes can induce charges on the surface of the top hBN, being consistent to the results on PL and transport properties of samples prepared with the nano-squeezing method.

To suppress the effect from induced charged impurities, we put a few-layer graphene on top of a hBN/MoSe2/hBN before a squeezing process. The graphene capping layer can effectively screen Coulomb interaction arising from the charged impurities induced on the topmost surface and thereby leads to ultra-flat monolayer $MoSe_2$ free from the effect of charged impurities[45, 46]. Figure 5a shows an optical microscope image of the hBN/MoSe2/hBN sample with a few-layer graphene capping layer, where a bottom graphene was additionally put under the hBN-encapsulated $MoSe_2$ to screen Coulomb interaction from charged impurity in the underlying SiO2/Si substrate; the region surrounded by the red box in the image was squeezed with applied force of 1.5 μN (Fig. 5b). As shown in Fig. 5c, the PL spectrum measured at 10 K clearly show very low intensity of the low-energy peak at 1.61 eV, meaning that the few-layer graphene capping layer successfully screens Coulomb potential arising from the charged impurities; the very low intensity also means minimal unintentional doping into the encapsulated monolayer $MoSe_2$. Residual linewidth of exciton emission determined through fitting with Voigt function is small, 2.2 meV (Fig. S5), which is comparable to the intrinsic PL linewidth reported[24, 47, 48].

Reflecting the high quality of the sample prepared with the modified nano-squeezing method, efficient exciton-exciton collisions were observed. Figure 6a shows an excitation power dependence of PL spectra measured with 633 nm continuous wave (CW) laser. As clearly seen, a peak, whose intensity drastically increases as excitation laser power increases, emerges at ~ 1.62 eV in addition to excitonic peak at ~ 1.645 eV; the low-energy peak is denoted by X′. To address the origin of X′, we plotted $I_{ex}$ and intensities of X′ with respect to photon flux ($\Phi_{ph}$). As shown in Fig. 6b, while $I_{ex}$ is proportional to $\Phi_{ph}$ as expected, intensities of X′ are proportional to $\Phi_{ph}^{1.8}$, which strongly suggest that X′ originates from an excited state formed through two-body collisions of excitons. There are two possible origins for the excited state, biexcitons[39, 41, 48-50] and trions. Binding energy alone is not sufficient to unambiguously assign X′ because of small difference in binding energy between biexcitons and trions in MoSe$_2$: trion: 27.7 ~ 31 meV and biexciton: 17.7 ~ 23 meV[39, 47, 58-60]. To address the origin of X′, we investigated excitation-power dependence on peak position of X′ (Fig. S9). If X′ arises from trions, photo-induced carriers, which are generated through exciton-exciton annihilation processes, should exist and the number of the photo-induced carriers should increase as $\Phi_{ph}$ increases. In this case, position of X′ should show energy shift to the red side due to the increase of quasi-Fermi level[61]. As clearly seen in Fig. S9, this is totally inconsistent with the observed blue shift, suggesting X′ originates from biexcitons. Moreover, CCD images of PL spectra measured with different $\Phi_{ph}$ (Fig. S10) shows no enhancement in diffusion, which enhancement is expected due to formation of hot excitons through exciton-exciton annihilations. We therefore conclude that X′ can be assigned as biexciton emission. Note that this peak does not originate from the formation of excited states of excitons (2s or 3s state)[7, 48], which is another possible process when two excitons collide, because the energy difference between the low-energy peak and the exciton peak is less than that of the excited states and the ground state.

In the sample prepared with the modified nano-squeezing method, exciton-exciton collision is greatly enhanced. As seen in the Fig. 6a, generation of biexcitons occurs even at small $\Phi_{ph}$ of 2.3 × 10$^{19}$ cm$^{-2}$ s$^{-1}$ whereas the biexciton formation has not been seen in samples prepared without a graphene capping layer even with much higher $\Phi_{ph}$ of 5.1 × 10$^{21}$ cm$^{-2}$ s$^{-1}$ and even with pulsed excitations (pulse duration ~ 60 ps) with photon density of 5.1 × 10$^{14}$ cm$^{-2}$/pulse (Fig. S2b). To make sure the generation of biexcitons is reasonable, we compared average exciton-exciton distance $d_{ex}$ with exciton diffusion length $L_{ex}$. The $d_{ex}$ can be calculated as $d_{ex} = 1/\sqrt{\alpha \cdot \tau_{ex} \cdot \Phi_{ph}}$, where $\alpha$ and $\tau_{ex}$ represent absorptance and exciton lifetime, respectively. Using values of $\alpha$ = 0.07[31], $\tau_{ex}$ = 47 ps and $\Phi_{ph}$ = 2.3 × 10$^{19}$ cm$^{-2}$ s$^{-1}$, we obtained $d_{ex}$ = 1.1 μm; lifetime of excitons is measured with time-correlated single photon counting (TSCPC) method (Fig. S8). The $L_{ex}$ can be calculated with a following equation, $L_{ex} = 2\sqrt{D_{ex}\tau_{ex}}$, where $D_{ex}$ is diffusion coefficient. The $D_{ex}$ is determined with the Einstein relation, $D_{ex} \sim k_B T/M_{ex}\gamma$, where $k_B$, $M_{ex}$ and $\gamma$ correspond to Boltzmann's constant, exciton translational mass (~ 0.3 $m_0$)[51] and homogeneous linewidth of exciton emission (half-width at half-maximum of Lorentzian linewidth obtained through fitting with Voigt function)[52]. In this case, the $\gamma$ is only 0.5 meV at 10 K, corresponding to momentum relaxation time of 2.3 ps (Fig. S11). Using the determined value of $D_{ex}$, 11 cm$^2$ s$^{-1}$, the $L_{ex}$ is calculated as 0.5 μm, being comparable with the $d_{ex}$ calculated with the smallest $\Phi_{ph}$ used in this experiment. This indicates that diffusion alone can cause inter-exciton collisions to form biexcitons in the present high-quality sample. In addition, inter-exciton attractive interactions, such as phonon-mediated effective interaction[56, 57], may drive drift of excitons, which also can enhance inter-exciton collisions in the high-quality sample.

In conclusion, we have successfully developed a sample preparation technique, the modified nano-squeezing method, to surely fabricate high-quality vdW heterostructures. The key idea in this technique is squeezing bubbles and impurities through a few-layer graphene capping layer to have ultra-flat heterostructures free from random electrostatic potential arising from charged impurities. Without a graphene capping layer, a sweeping with an AFM tip induces applied-force-dependent charges of $\Delta n \sim 2 \times 10^{12}$ cm$^{-2}$ μN$^{-1}$, leading to strong PL intensity of trions and level off of carrier mobility at low temperature. A hBN/MoSe$_2$/hBN prepared with the modified nano-squeezing method shows a strong excitonic emission with negligible trion peak, and the residual linewidth of the excitonic peak is only 2.2 meV. Furthermore, in this high-quality sample, we found that formation of biexciton occurs even at very low excitation power ($\Phi_{ph} \sim 2.3 \times 10^{19}$ cm$^{-2}$ s$^{-1}$) due to the enhanced collisions between excitons arising from long exciton diffusion length. This work has established a new avenue for exploration of physics of 2D materials based on ultra-flat vdW heterostructures free from random electrostatic potential.

Experimental

Fabrication of vdW heterostructures

Monolayer MoSe$_2$, graphene and hBN flakes used to fabricate vdW heterostructures were mechanically exfoliated from bulk crystals and deposited onto SiO$_2$/Si substrates. Firstly, a hBN flake was picked up with a polymer stamp[53, 54], and then other flakes were consecutively picked up with the hBN flake on the polymer stamp. Note that we heated substrates up to 110 ~ 130 °C to make contact between flakes better during pick-up processes. Finally, the polymer stamp with vdW heterostructure on top was transferred onto another hBN flake on a SiO$_2$/Si substrate. Polymer on the vdW heterostructure was removed by soaking in chloroform overnight. In the case of a sample with a graphene capping layer, instead of a hBN flake, a few-layer graphene flake was picked up first and other flakes were consecutively picked up. To remove bubbles and impurities encapsulated between layers, we applied force by sweeping surface of samples with an AFM-tip working under contact mode (BRUKER Inc., Dimension FastScan). Before sweeping sample surface, applied force was calibrated by bare surface of SiO$_2$/Si substrates using the torsional Sader method[55]. Typical scan lines and scan rate in the sweeping process were 512 – 2024 lines and 0.8 – 1 Hz, respectively.

Electrostatic force microscopy (EFM) measurement

EFM images were obtained with a commercial AFM setup (Park SYSTEMS, NX10 EFM

operating at EFM mode) and a conductive AFM cantilever (Olympus, OMCL-AC240TM Pt coated). We made electrical contact between substrates and the sample stage by Ag paste. All EFM images were taken under the following experimental conditions: 512 lines, 25×25 μm$^2$, Scan rate: 0.3 – 0.05 Hz, sample bias voltage: 0 V and dark environment.

Optical measurements

Optical responses were obtained by a home-made microspectroscopy system. We used a HeNe CW laser (Thorlabs, HNL050L) and a pulsed super-continuum white laser (SuperK EXTREME, NKT Photonic, 40 MHz) for sample excitations. The super-continuum laser was monochromated by a spectrometer (Princeton Instruments, SP2150) to obtain desired wavelength. The laser beam was focused on a sample by a x50 objective lens with a correction ring (Nikon, CFI L Plan EPI CR, *NA* = 0.7). Samples were set on a sample stage cooled by flowing He liquid under vacuuming (KONTI-Cryostat-Micro, CryoVac) and sample temperature was monitored and controlled by a temperature controller (TIC 304-MA, CryoVac). The PL signal extracted by a long pass filter (Thorlabs, FELH series) was introduced to a spectrometer (IsoPlane 320, Princeton Instruments) and detected by a charge coupled device (PIXIS 1024B_eXcelon, Princeton Instruments). We measured time-resolved PL (TRPL) by time-correlated single photon counting (TCSPC) method with an avalanche photo detector (Becker & Hickl GmbH, ID-100-50-ULN). The obtained data were fitted by double exponential decay function ($I(t) = a_1 \exp(-t/\tau_1) + (1 - a_1)\exp(-t/\tau_2)$), where $\tau_1$, $\tau_2$ and $a_1$ are radiative lifetimes and a fitting parameter.

Acknowledgments

R. K. was supported by JSPS KAKENHI Grant numbers JP16H03825, JP16H00963, JP15K13283, JP25107002, and JST CREST Grant Number JPMJCR16F3. T. H. was supported by JSPS KAKENHI Grant number JP19J15359. M. O. was supported by JSPS KAKENHI Grant number JP19K15403. K.W. and T.T. acknowledge support from the Elemental Strategy Initiative conducted by the MEXT, Japan, Grant Number JPMXP0112101001, JSPS KAKENHI Grant Numbers JP20H00354 and the CREST(JPMJCR15F3), JST.

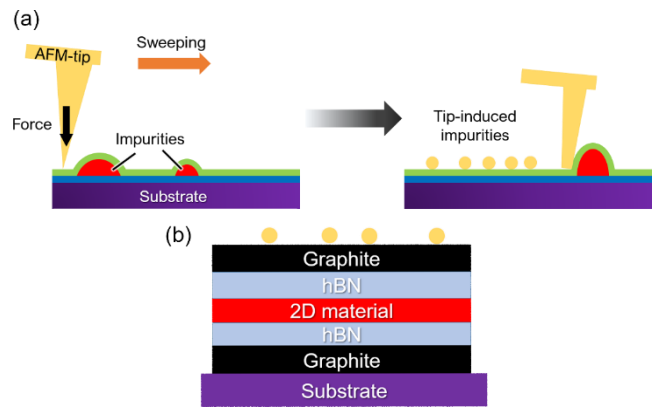

Figure 1 (a) A schematic image showing the nano-squeezing process by an AFM tip sweeping. Impurities encapsulated between layers can be squeezed out, but sweeping-induced impurities attach to the surface. (b) A schematic image of a hBN-encapsulated 2D material with a few-layer graphene capping layer, which can screen random Coulomb potential arising from the sweeping-induced impurities. A graphite bottom layer is also put to screen Coulomb potential arising from charged impurities locate at the underlying $SiO_2$/Si substrate.

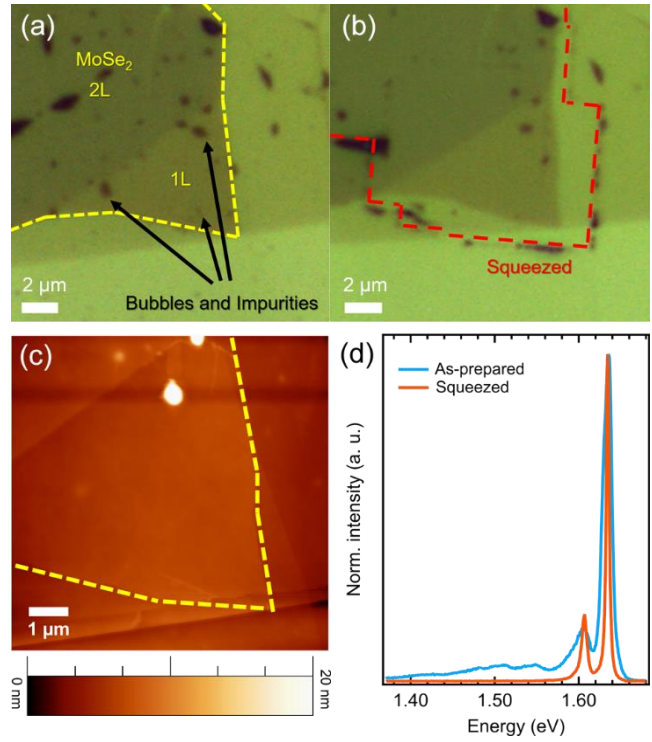

Figure 2 (a), (b) and (c) Optical microscope and AFM images of a hBN/MoSe$_2$/hBN sample before and after a squeezing process with applied force of 0.8 μN. Black dots in the optical images correspond to impurities/bubbles encapsulated between the MoSe$_2$ and hBN flakes. (d) Typical PL spectra of the sample before (blue) and after (orange) the squeezing process. Both spectra were measured at 10 K with excitation wavelength of 550 nm. After the squeezing process, reduction of linewidth and background are seen.

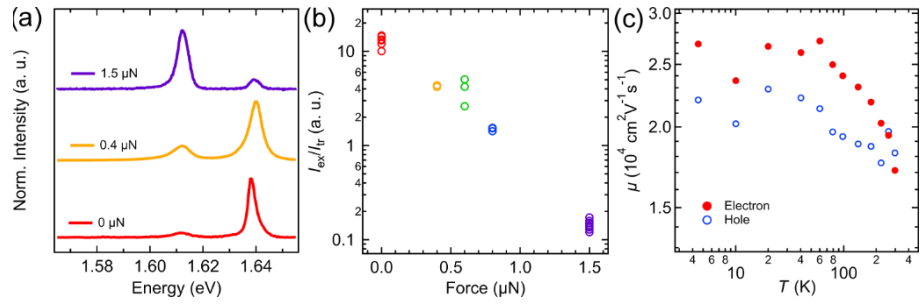

Figure 3 (a) PL spectra of hBN/MoSe$_2$/hBN samples measured at 10 K. Purple, orange and red curve correspond to spectra of samples squeezed at 0, 0.4 and 1.5 µN, respectively. (b) Applied-force dependence of ratio of peak area of exciton and trion emission. (c) Temperature dependence of Hall mobility of electrons and holes in a hBN/graphene/hBN Hall bar. Saturation in Hall mobilities, which is caused by the charged impurity scatterings, are seen at low-temperature region.

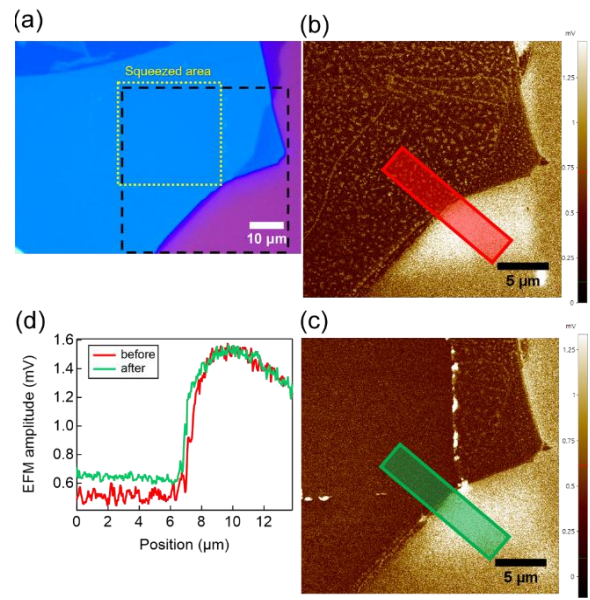

Figure 4 (a) An optical image of a hBN/hBN sample fabricated with the dry transfer method. A region surrounded by the yellow box is swept with applied force of 1.5 µN. (b) (c) EFM images before and after the squeezing process. (d) Line profiles of the EFM images along the long sides of red and green rectangles. The profiles are averaged using the regions surrounded by the rectangles. To cancel out possible shift arising from difference between AFM tips, EFM amplitude is shifted to make the EFM amplitudes at $SiO_2$ surface same.

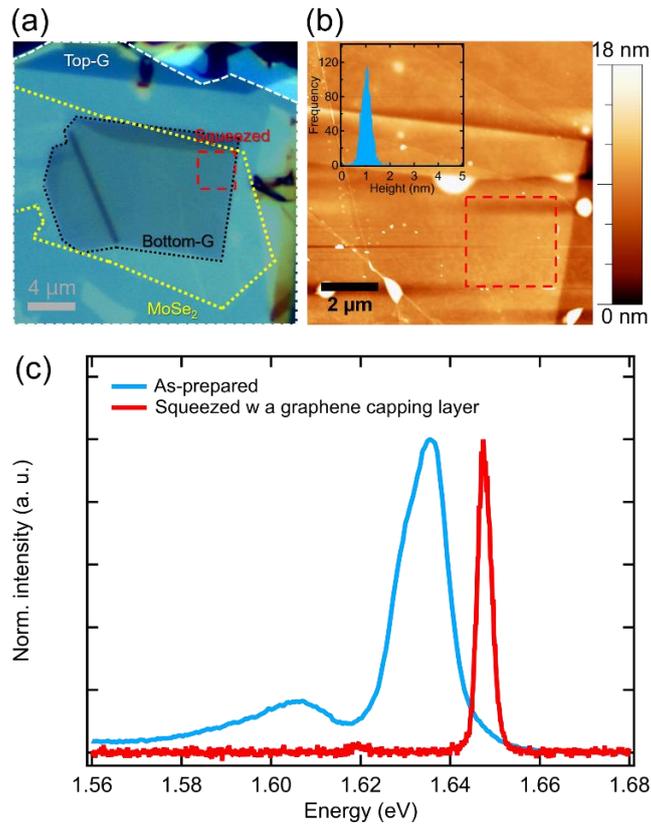

Figure 5 (a) An optical microscope image of a hBN/MoSe$_2$/hBN sample prepared with the modified nano-squeezing method. Edges of the few-layer graphene capping layer and the bottom graphite are shown as white and black dotted lines, respectively. (b) An AFM image of the squeezed sample. The red dashed square shows the region squeezed with applied force of 1.5 μN. (inset) A height frequency distribution obtained at the squeezed region. RMS surface roughness reduces from 0.76 to 0.31 nm after the squeezing process. (c) PL spectra of hBN/MoSe$_2$/hBN samples, an as-prepared sample and the sample prepared with the modified nano-squeezing method, measured at 10 K; excitation wavelength and photon flux are 633 nm and $\Phi_{ph} \sim 2.3 \times 10^{19}$ cm$^{-2}$s$^{-1}$, respectively.

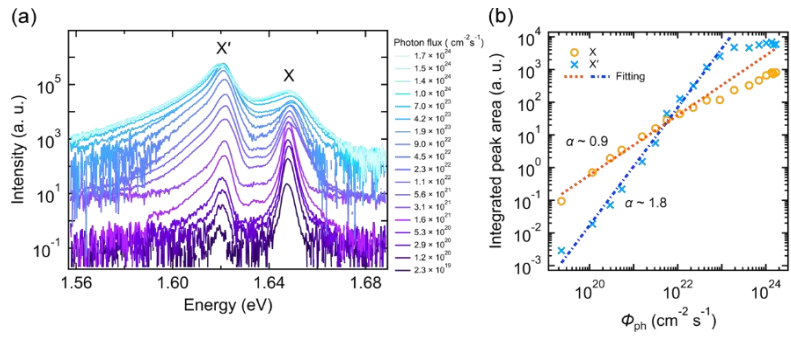

Figure 6 (a) Photon flux dependence of PL spectra measured at 10 K. The sample prepared with the modified nano-squeezing method is excited at 633 nm HeNe CW laser. (b) log-log plots of integrated peak PL intensity versus photon flux, $\Phi_{ph}$. Orange and blue points correspond to integrated peak intensity of exciton and trion emission, respectively. These plots can be fitted with exponent α of ~ 0.9 and ~ 1.8. Even in the smallest $\Phi_{ph}$, generation of biexcitons is clearly seen.

## S1 Optical and AFM images of hBN/MoSe$_2$/hBN heterostructures

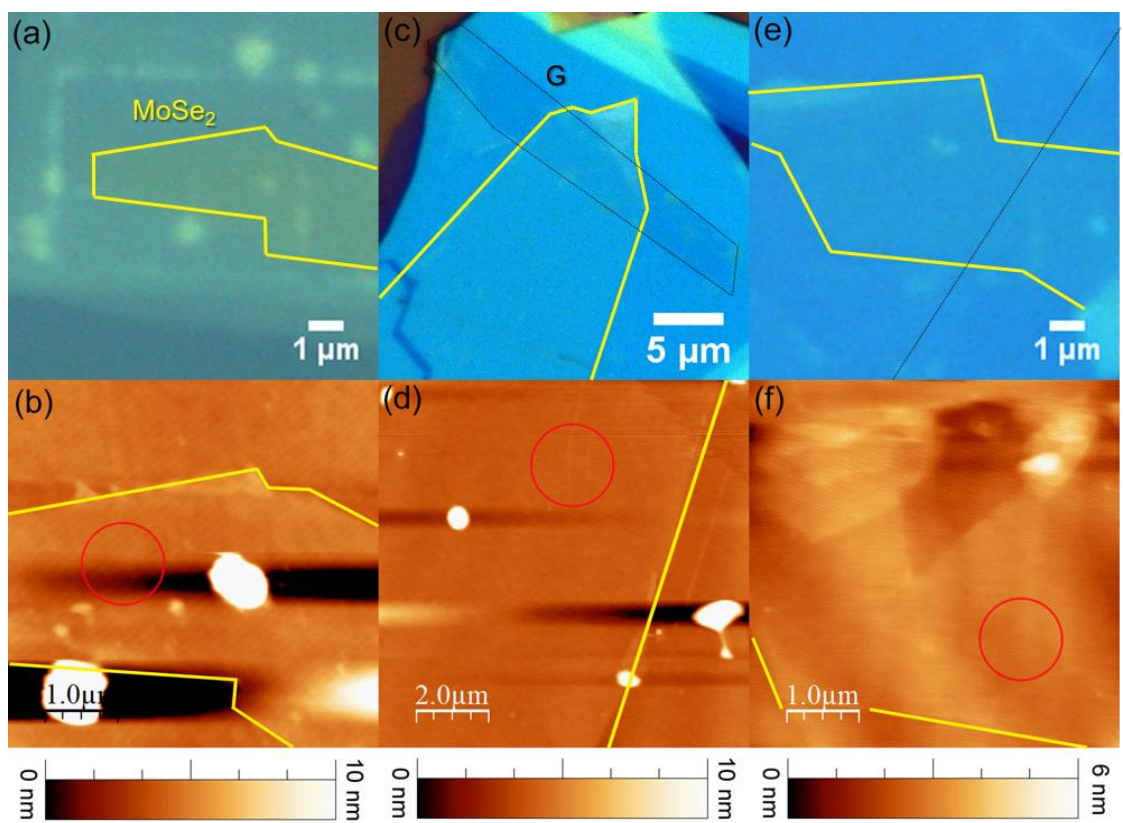

Figure S1 Optical and AFM images of fabricated hBN/MoSe$_2$/hBN heterostructures squeezed with applied force of (a) (b): 0.4 µN, (c) (d): 0.6 µN and (e) (f): 1.5 µN. Yellow lines show edges of monolayer MoSe$_2$ flakes encapsulated between hBN flakes. PL spectra were obtained at around red circles, where no bubbles and impurities were seen. Black lines in Fig. 1S(c) and (e) correspond to edges of graphene flakes for electrical contacts.

# S2 PL spectra of the hBN/MoSe$_2$/hBN heterostructure squeezed at 1.5 μN measured at 10 K

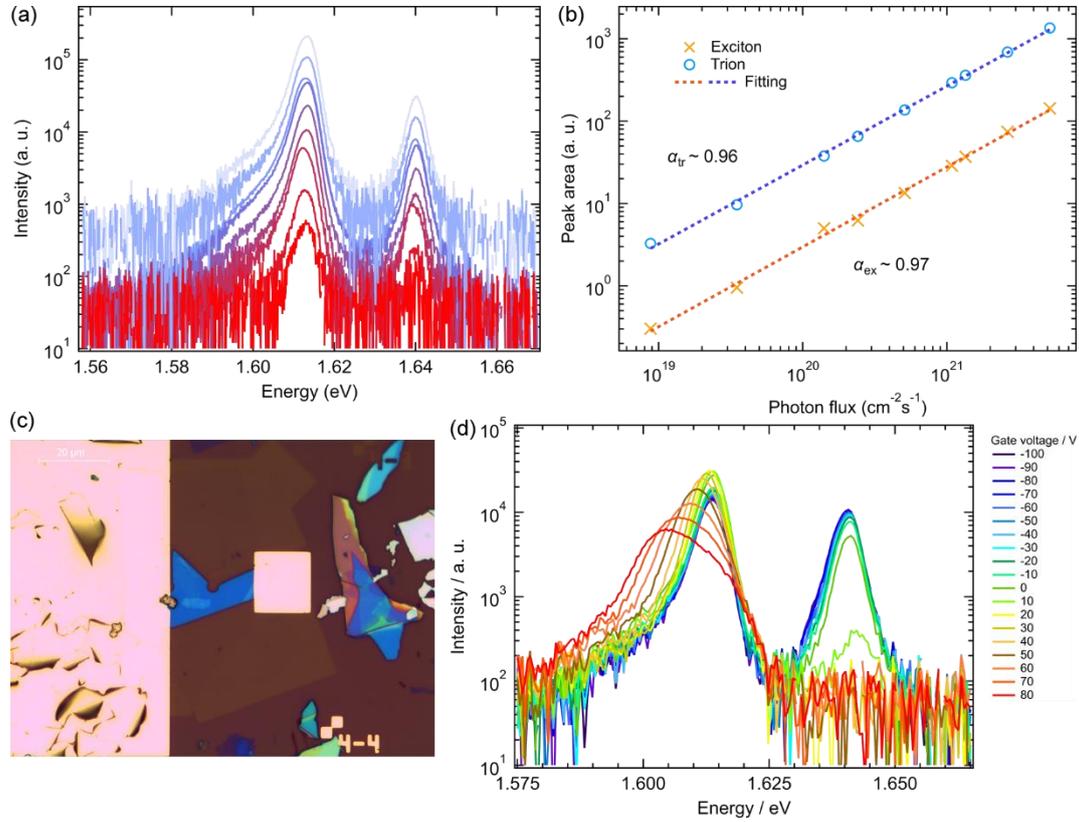

Figure S2 (a) An excitation power dependence of PL spectra of the hBN/MoSe$_2$/hBN heterostructure squeezed at 1.5 μN. A 633 nm HeNe CW laser was used for the excitation. Two peaks at ~ 1.61 eV and ~ 1.64 eV, derived from emission from exciton and trion, were seen in all spectra. (b) An excitation power dependence of the peak areas of the exciton and trion peaks. Orange and blue points correspond to peak areas of exciton and trion, respectively. As shown in the figure, peak area increases linearly with excitation power (exponent $\alpha$ ~ 0.97 and 0.96), meaning the low energy peaks at 1.61 eV can be assigned to trions even at $\Phi_{ph}$ ~ 5.2 × 10$^{21}$ cm$^2$ V$^{-1}$ s$^{-1}$. (c) An optical image of the fabricated hBN/MoSe$_2$/hBN device. The electrical contacts were made through graphene on the MoSe$_2$. (d) A gate-voltage dependence on PL spectra measured at 10 K. With increasing the gate voltage, the intensity of exciton (trion) decreases (increase).

## S3 Comparing to electrical and tip-induced carrier doping

In our result, applied-force dependence on $I_{ex}/I_{tr}$ can phenomenologically be fitted by single-exponential decay ($y = A\exp(-\alpha x)$) (Fig. S3(a)) with parameters of A = 14.89 and $\alpha$ = 3.05 × $10^{-3}$ µN$^{-1}$. To estimate the amounts of tip-induced carriers, we compere the obtained applied-force dependence to a gate-voltage dependence on $I_{ex}/I_{tr}$. For this purpose, we made electrical contact (Au/Cr 80/10 nm) to the sample squeezed at 0.6 µN, where the contact was made through a graphene flake placed on top of MoSe$_2$ (Fig. S3(b)). Fig. S3(c) shows a gate-voltage dependence on PL spectra measured at 10 K. As clearly seen, trion intensity increases as gate voltage increases, which results from the injection of elections into MoSe$_2$. The number of carriers injected can be calculated based on the capacitance model, $\Delta n = C_{tot} V_g / e$; $C_{tot}$ corresponds to the total capacitance per unit area of insulators, $1/C_{tot} = 1/C_{SiO_2} + 1/C_{hBN}$. Each capacitance can be calculated by $C = \varepsilon_r \varepsilon_0 / d$, where $\varepsilon_r$, $\varepsilon_0$ and $d$ represent relative dielectric constant of materials, dielectric constant of vacuum and thickness of the insulator layer (hBN: 5.01[1] and 40 nm / SiO$_2$: 3.8 and 270 nm), respectively. Fig. S3d represents peak areas of exciton and trion peaks against $\Delta n$, showing rapid decrease (increase) of $I_{ex}$ ($I_{tr}$) upon increase of $\Delta n$. $I_{ex}/I_{tr}$ thus shows steep exponential decrease depending on $\Delta n$ as shown in Fig. S3(e). The exponential decrease at the positive side of $V_g$ can be well fitted by single-exponential decay with parameters of A = 2.38 and $\alpha$ = 1.80 × $10^{-12}$ cm$^2$.

As shown above, both plots of applied-force and gate-voltage dependence can be phenomenologically fitted by the single-exponential decay. This indicates that the number of carriers is proportional to applied forces, which is consistent to the well-known linear relation between applied force and abrasion loss. The applied forces, therefore, can be converted to carrier density with a conversion factor of ~ 2 × $10^{12}$ cm$^{-2}$µN$^{-1}$. The open circles and black dashed line in Fig. S3e correspond to the data shown in Fig. S3(a), whose horizontal axis is converted with the conversion factor.

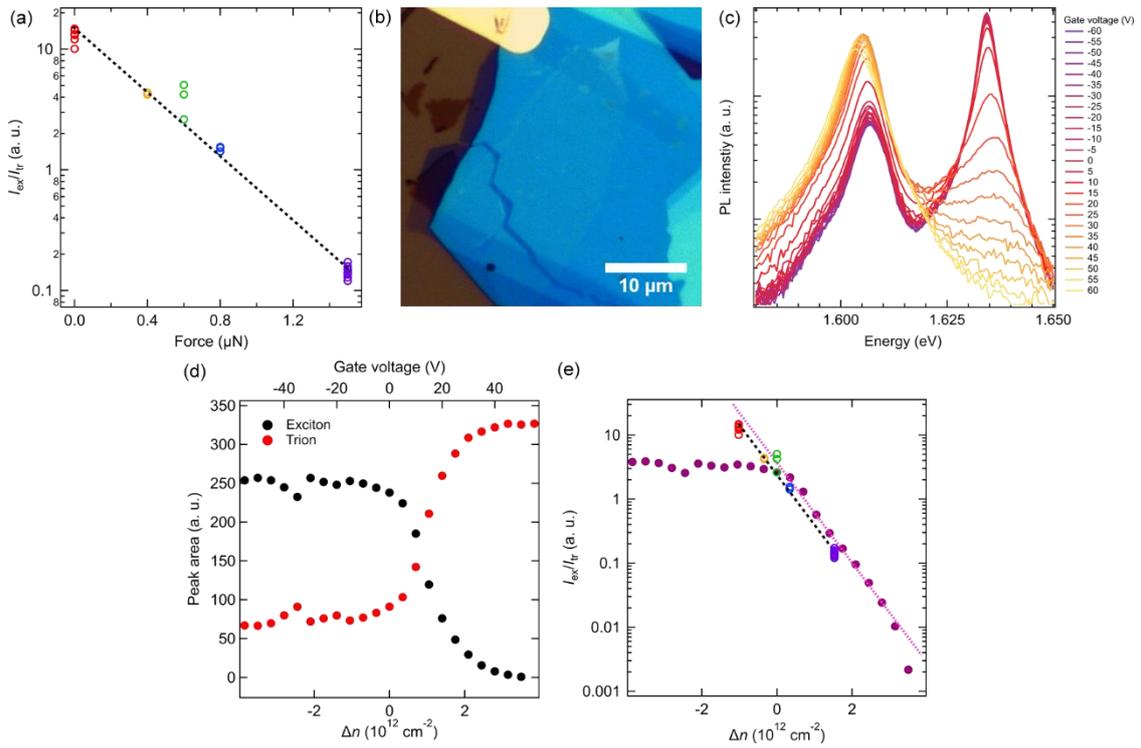

Figure S3 (a) An applied-force dependence on $I_{ex}/I_{tr}$ and a fitting line assuming single-exponential decay (dashed line). (b) An optical image of a fabricated hBN/MoSe$_2$/hBN heterostructure. (c) PL spectra measured with gate voltages from -60 V to 60 V and excitation wavelength of 550 nm. (d) A gate voltage (carrier density) dependence of peak areas of exciton and trion peaks. (e) Δn dependence of $I_{ex}/I_{tr}$, where a purple dashed line corresponds to the fitting result.

## S4 A graphene hall bar

A hBN/graphene/hBN heterostructure was fabricated by the dry transfer method. Bubbles and impurities were squeezed out with applied force of 1.8 µN (Fig. S4(a) and (b)). Although bubbles and impurities were not fully removed, the region surrounded by the white dotted box was atomically flat (RMS: 0.18 nm). We selected the flat region for fabrication of a graphene hall bar (Fig. S4(c)); we used the standard microfabrication techniques, including electron beam lithography, reactive ion etching (RIE), and vacuum metal deposition. Fig. S4(d) shows transfer characteristics measured at temperature range from 4.5 to 300 K. Maximum resistivities appears at $V_g > 0$, meaning that extrinsic doping exists even with ultra-flat graphene encapsulated by hBN flakes.

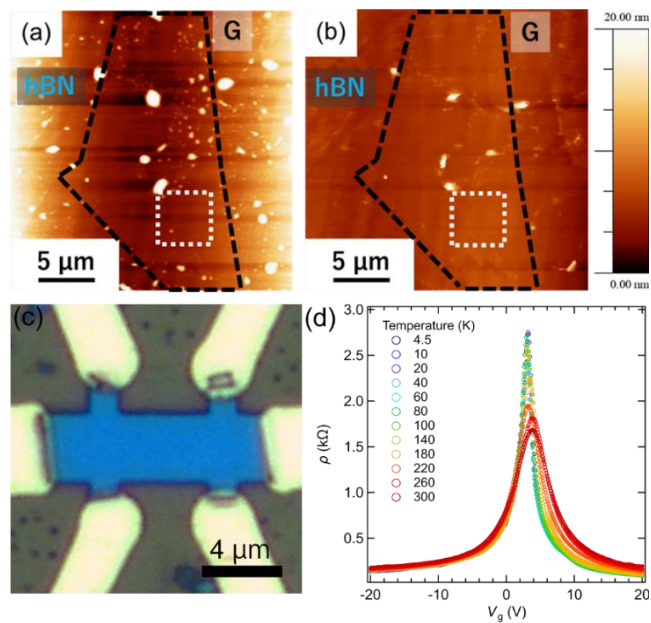

Figure S4 (a) (b) An AFM image before and after a squeezing process with applied force of 1.8 µN. In the white dotted box, RMS reduces from 0.51 to 0.18 nm after the squeezing process. (c) An optical image of the fabricated hBN/graphene/hBN hall bar on a 270 nm $SiO_2$/Si substrate. (d) Transfer characteristics measured at temperature range from 4.5 to 300 K.

## S5 Estimation of residual linewidth at 0 K

To estimate residual linewidth at 0 K, temperature dependence of excitonic linewidth is evaluated. Fig. S5 represents temperature dependence of PL spectra; all spectra were measured with 633 nm pulsed laser excitation with photon density of $2.5 \times 10^{10}$ cm$^{-2}$/pulse. Slight blue shift is seen in these PL spectra, which arises from enlargement of bandgap caused by electron-phonon coupling. By fitting these spectra, we extracted linewidth of excitonic emission (FWHM). Fig. S5b shows temperature dependence in the FWHM. The temperature dependence was fitted with the equation shown below; the first, send and third term correspond to the residual linewidth, the contribution of acoustic phonon and optical phonon scattering, respectively.

$$\gamma = \gamma_0 + c_1 T + \frac{c_2}{\exp\left(\frac{\Omega}{kT}\right) - 1}$$

The $c_1$, $c_2$, $\Omega$, $k$ and $T$ represents a parameter for the acoustic phonon contribution, the optical phonon contribution, typical energy of optical phonon, Boltzmann constant and temperature. The fitting with the equation yielded $\gamma_0 = 2.2 \pm 0.2$ meV, $c_1 = 31 \pm 6$ µeV K$^{-1}$, $c_2 = 50 \pm 33$ meV and $\Omega = 30 \pm 12$ meV. At low temperature region, the third term is almost negligible, and the temperature dependence of linewidth is dominated by the 2$^{nd}$ term.

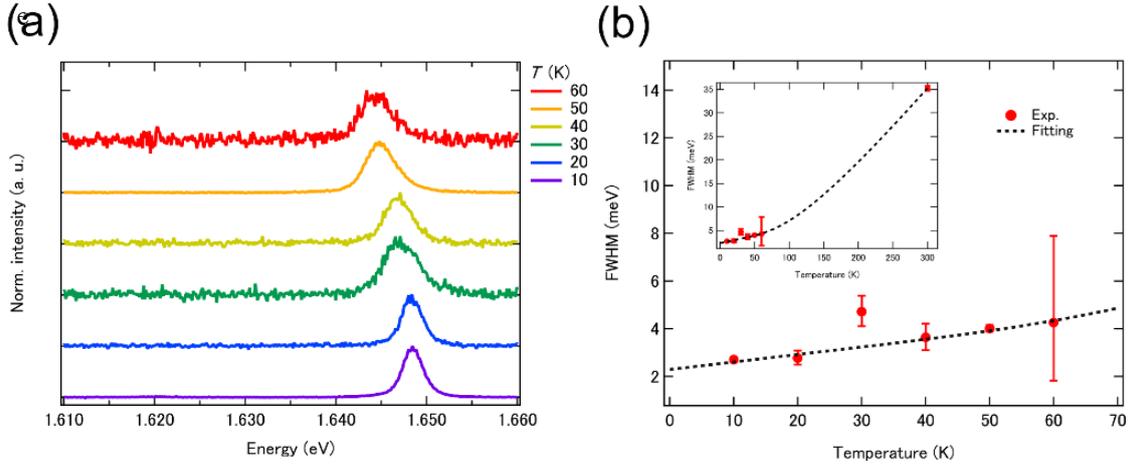

Figure S5 Temperature dependence of PL spectra ranging from 10 K to 60 K. All spectra are obtained with a 633 nm pulsed laser excitation: photon density is ~ $2.5 \times 10^{10}$ cm$^{-2}$/pulse. (b) Temperature dependence of FWHM extracted. Red dots and black dashed curve show the extracted FWHM and fitted by the above equation. (inset) this fitting is executed ranging from 10 – 60 K and 300 K.

## S6 A log-log plot between integrated peak areas of exciton and biexciton

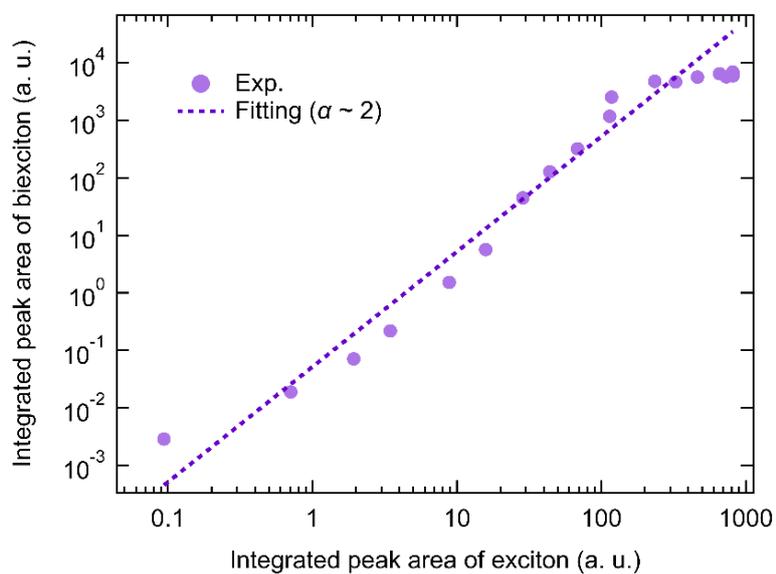

Figure S6 Integrated peak areas of biexciton dependent on those of exciton. The area of exciton $I_{ex}$ is proportional to the area of biexciton $I_{bi}^2$, which clearly demonstrates that the low-energy peak arises from excited states formed through two-body collisions formation of excitons.

**S7 Excitation power dependence of PL spectra excited with a pulsed laser**

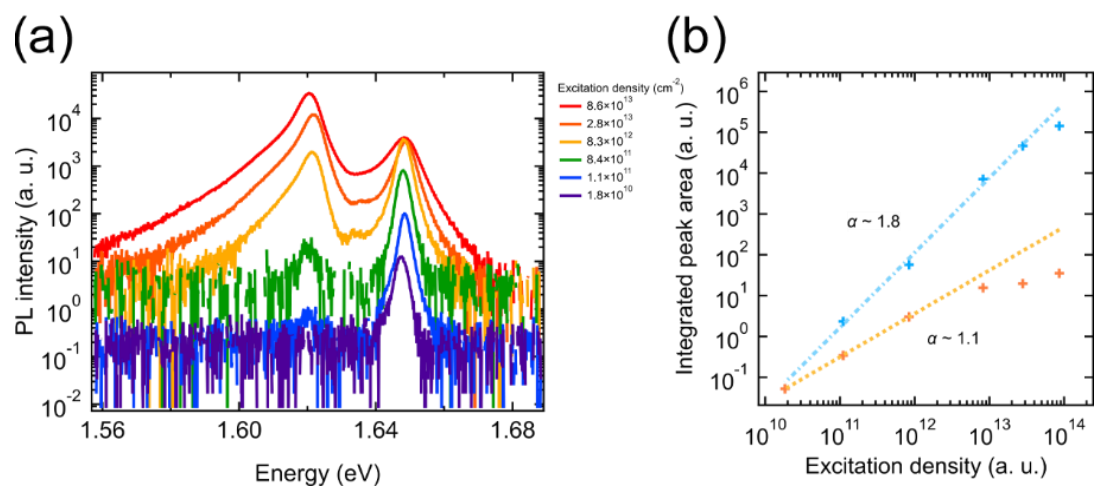

Figure S7 (a) Excitation power dependence of PL spectra measured at 10 K with excitation wavelength of 633 nm. We used a pulsed laser with photon density ranging from 1.8 × $10^{10}$ to 8.6 × $10^{13}$ cm$^{-2}$/pulse. (b) Excitation photon density dependence of integrated peak area. As seen in the excitation power dependence measured with the CW laser excitation, the low-energy peak shows superlinear relation with exponent of 1.8.

## S8 Radiative lifetime of excitons measured by TCSPC method

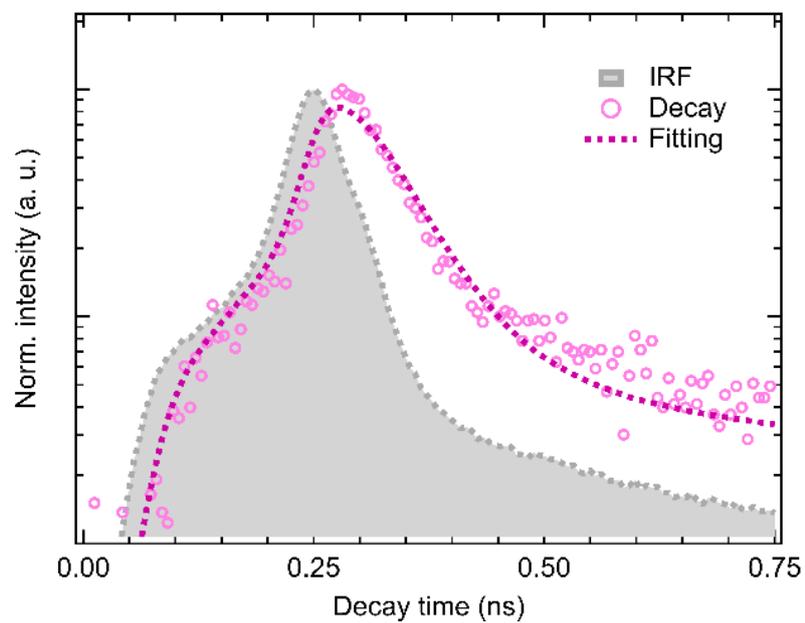

Figure S8 PL decay at 10 K measured by TCSPC method. Gray area and pink dots correspond to instrument response function (IRF) and PL decay, respectively. We fitted the PL decay with a double exponential decay function (dashed line).

.

## S9 Photon flux dependence of peak positions

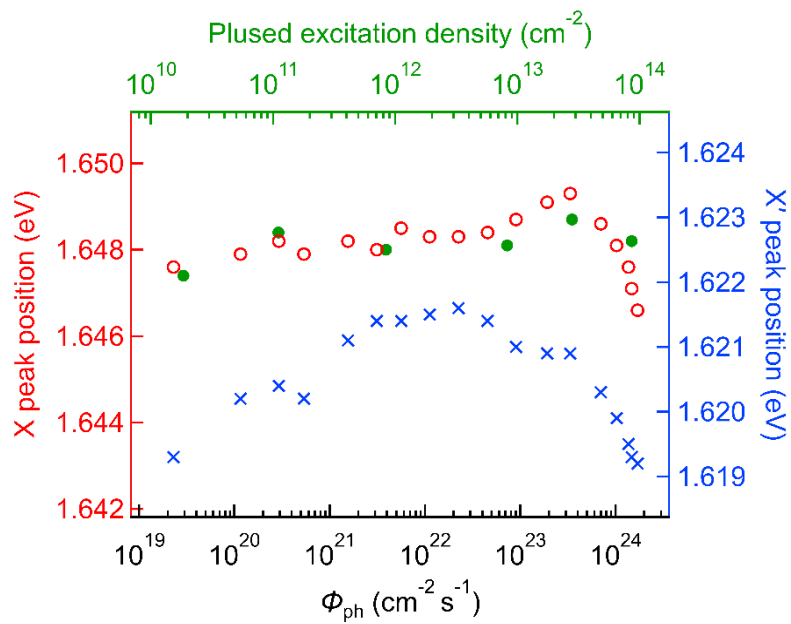

Figure S9 Excitation photon flux dependence of peak positions. Green filled circles correspond to positions of PL peaks arising from excitons measured with pulsed excitations. Red open circles and blue crosses correspond to peak positions of excitons and the low-energy peak measured with CW excitations. Both peaks show slight blue shift as excitation photo flux become larger. Excitonic peak shifts excited by the pulsed laser are similar trend to that of CW excitation, meaning heating effect by CW excitation is less.

## S10 CCD images of PL spectra

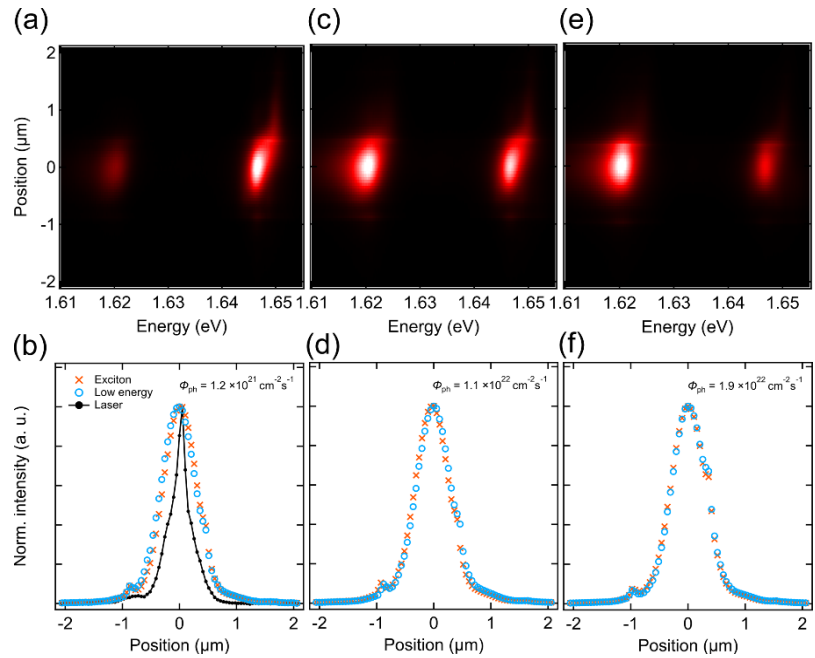

Figure S10 CCD images of PL spectra measured at 10 K with excitation photon flux of (a) $1.2 \times 10^{21}$ cm$^{-2}$s$^{-1}$, (c) $1.1 \times 10^{22}$ cm$^{-2}$s$^{-1}$, (e) $1.9 \times 10^{22}$ cm$^{-2}$s$^{-1}$, respectively. PL intensities are normalized by the maximum value. (b), (d) and (f) show line profiles of exciton and the low-energy peaks. The black line and dots in Fig. S10b corresponds to a line profile of the excitation laser spot.

## S11 Extracting homogeneous linewidth by Voigt function

To extract homogeneous linewidth of exciton emissions, we used 5 different spectra of the same graphene-capping hBN/MoSe$_2$/hBN sample measured with low excitation power of $2.3 \times 10^{17}$ ~ $1.6 \times 10^{19}$ cm$^{-2}$s$^{-1}$. Fitting with Voigt function gives Gaussian width and Lorentzian width, which correspond to inhomogeneous broadening and phase relaxation, respectively. Assuming that Gaussian width are the same in all spectrum, we used averaged Gaussian width, 2.9 meV, obtained through fitting of the 5 spectra. Fig. S6 shows a PL spectrum of the graphene-capping sample and the peak is fitted by Voigt function with parameters of Lorentzian width of 0.5 meV.

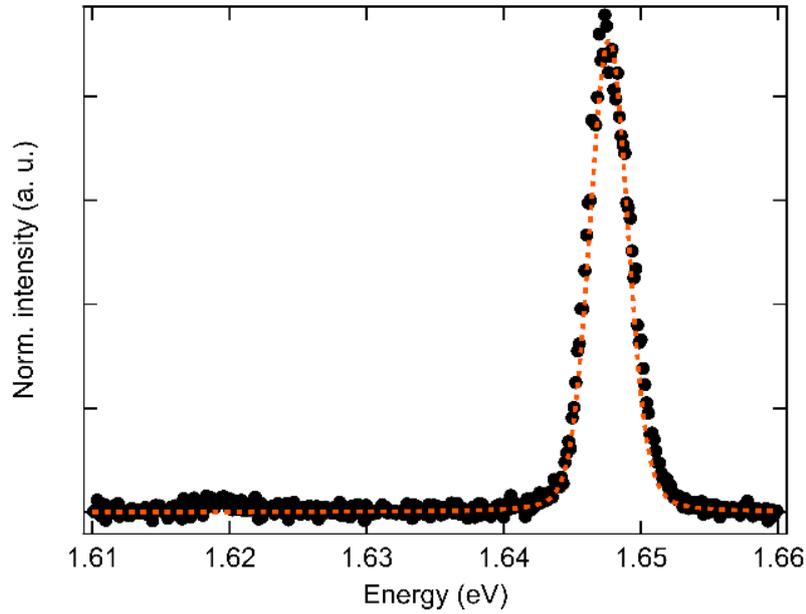

Figure S11 PL spectra at 10 K (black dots) and the fitting result with parameters of 2.9 meV (Gaussian) and 0.5 meV (Lorentzian). The sample was excited at $\Phi_{ph}$ ~ $2.3 \times 10^{19}$ cm$^2$ V$^{-1}$ s$^{-1}$.